\documentclass[]{raa}            
\usepackage{graphicx,times}
\usepackage{natbib}

\providecommand{\bm}[1]{\mbox{\boldmath$#1$\unboldmath}}

\def\jn{\mbox{PKS 0138$-$097}}
\usepackage{amssymb}
\usepackage{amsfonts}
\usepackage{txfonts}
\usepackage{natbib}

\newcommand{\zsh}[1]{{\color{black} #1}}

\bibpunct{(}{)}{;}{a}{}{,}

\begin{document}

\title{Discovery of A Variable Broad Absorption Line in the BL Lac object PKS 0138$-$097}
\volnopage{ {\bf 2011} Vol.\ {\bf 000} No. {\bf XX}, 000--000}
   \setcounter{page}{1}
\author{ Shaohua Zhang\inst{1,2}
Huiyuan Wang\inst{1}, Hongyan Zhou\inst{1,2}, Tinggui Wang\inst{1},
Peng Jiang\inst{1} }
\institute{ Key Laboratory for Researches in Galaxies and Cosmology, Department of Astronomy,
University of Sciences and Technology of China, Chinese Academy of Sciences, Hefei, Anhui 230026, China;
zsh@mail.ustc.edu.cn, whywang@mail.ustc.edu.cn, mtzhou@ustc.edu.cn
\and Polar Research Institute of China, Jinqiao Rd. 451, Shanghai, 200136, China }

\abstract{ We report the discovery of a Broad Absorption Line (BAL) of $\sim
10^{4}~\rm km~s^{-1}$ in width in the previously known BL Lac \zsh{object} PKS 0138$-$097,
which we tentatively identified as  \zsh{a} Mg~II BAL. This is the first
detection of a BAL\zsh{, which is sometimes seen in powerful quasars
with high accretion rates,} in a BL Lac \zsh{object}.
\zsh{The BAL was clearly detected in its spectra of two epochs
at a  high luminosity state taken in the Sloan Digital Sky Survey (SDSS),
while it disappeared in three SDSS spectra taken at a low luminosity state.}
The \zsh{BAL and its} variability pattern was also \zsh{found} in \zsh{its}
historical multi-epoch spectra in the literature, but has been
overlooked \zsh{previously}. \zsh{In its high resolution radio maps,}
PKS 0138$-$097 \zsh{shows} a core plus \zsh{an} one-sided parsec-scale jet.
The BAL variability can be interpreted as follows: The optical emission is
dominated by the core in a high state and by the jet in a low state, \zsh{and the BAL material is}
located between the core and jet so that \zsh{the} BAL appears only when the
core is shining. \zsh{Our discovery suggests that outflows may also be produced in
active galactic nuclei at a low accreting state.}
}

   \authorrunning{ZHANG et al. }            
   \titlerunning{A Variable BAL BL Lac}  
\maketitle

\section{Introduction}
\label{sec_intro}
Evidence accumulated  in the
past decade points \zsh{to} that feedback from Active Galactic Nuclei
(AGNs) plays a crucial role in galaxy formation and evolution.
The accretion onto Super-Massive Black Holes (SMBHs) can release \zsh{a}
large amount of radiative and kinetic energy to their surroundings.
This may heat and expel the interstellar and intergalactic gas,
which serves as the common reservoir for both of SMBH growth and
star formation, and thus \zsh{regulates} the co-evolution of SMBHs and
their host galaxies (see Antonuccio-Delogu \& Silk 2010 for a recent
review). It is suggested that the kinetic energy output through jets
and outflows is at least as important as the radiative output in
most AGNs (Begelman 2004 and references therein).

The most energetic outflows \zsh{in AGNs are }  Broad Absorption Line (BAL)
systems\zsh{,} with width $\gtrsim 2000~\rm km~s^{-1}$ by definition. They
appear in the spectra of $\sim 15\%$ optically selected quasars, and
are often observed \zsh{as absorption by ions of}  C~IV, Si~IV, Al~III, and
Mg~II (Tolea et al. 2002; Hewett \& Foltz 2003; Reichard et
al. 2003; Trump et al. 2006; Gibson et al. 2009; Zhang et al. 2010). The velocity of
BAL outflows is typically a few thousand $\rm km~s^{-1}$, and \zsh{in} the
extreme case, can reach  \zsh{values} as large as $v\sim 0.2c$ (Foltz et
al. 1983).  It is now commonly accepted that BAL and non-BAL quasars
are physically the same\zsh{,} and a BAL Region (BALR) is present in all
quasars but with a small covering factor (CF). The dichotomy of BAL
and non-BAL quasars is attributed to an orientation effect: BALs are
observed in those quasars only when our line-of-sight passes
through the BALR. Such orientation models require a rather large
incline angle for BAL quasars with the BALR suggested to be an
equatorial wind driven from the accretion disk (e.g., Hines \& Wills
1995; Cohen et al. 1995; Goodrich \& Miller 1995; Murray et al.
1995; Elvis 2000). Evidence for the orientation models includes: the
similarity of \zsh{the} emission line spectra (Weymann et al. 1991) and the
spectral energy distribution (SED; e.g., Willott et al. 2003;
Gallagher et al. 2007) between BAL and non-BAL quasars; a small BALR
CF inferred from \zsh{the} emission line profiles (Korista et al. 1993) and
spectropolarimetric observations (e.g., Goodrich \& Miller 1995;
Cohen et al. 1995; Hines \& Wills 1995; Ogle et al. 1999; Schmidt \&
Hines 1999). However, it has been shown that BAL quasars \zsh{have}  on average
 higher accretion \zsh{rates} than non-BAL quasars (Boroson 2002;
Ganguly et al. 2007; Zhang et al. 2010). This is difficult to \zsh{understand}
 in the context of the simple inclination models. In order
to avoid the well-known inverse Compton catastrophe, Zhou et al.
(2006) constrained the inclination \zsh{angles} of the
outflows to be $\lesssim 20^{\circ}$ in six BAL quasars (see also
Ghosh \& Punsly 2007 for more examples), consistent with previous
radio morphology studies (Jiang \& Wang 2003; Brotherton et al.
2006). Follow-up XMM-Newton observations indicate that some of the
polar BAL quasars are transparent in the X-ray, contrary to what
observed in ``normal'' BAL quasars (Wang et al. 2008). The nature of
these mysterious polar BAL outflows is unclear, neither is their
relation with the relativistic jets.

Relativistic jets are believed to exist in radio-loud objects,
accounting for about 10\% of AGNs. \zsh{The} extreme \zsh{objects are} BL Lac
objects, which are characterized by weak emission lines,
rapid variability, high polarization, and a compact radio structure
(e.g., Stocke \& Rector 2000).
\zsh{BL Lac objects are extreme low-accretion  ratio ($\lesssim 10^{-2}$ of the Eddington ratio) AGNs
with the relativistic jet pointing toward the observer (e.g., Blandford \& Rees 1978; Urry \& Padovani 1995).
 Low accretion rates may lead the accretion flows to be advection-dominated (Narayan \& Yi 1995),
thus the accretion models in most BL Lac objects might be an advection-dominated accretion flow (ADAF) or
an ADAF in the inner region plus a standard thin disk (SD) in the outer region, i.e. ADAF+SD scenario
(Cao 2003). }

In this Letter, we report the discovery of a BAL in the classical low-frequency-peaked BL Lac object,
PKS 0138$-$097, from the 1 Jy sample (Stickel et al. 1991).
This is the first detection of a BAL in a BL Lac \zsh{object, which provides a unique opportunity
to study the possible relation between BAL outflows and jets.}
\zsh{This BL Lac object is multi-waveband variable (e.g., Stickel et al. 1993; Stocke \& Rector 1997;
Rector \& Stocke 2001; Barvainis et al. 2005) and highly polarized ($P=6\%-29\%$, Mead et al. 1990),
and its host galaxy has remained unresolved in the optical (e.g., Stickel et al. 1993;
Heidt et al. 1996; O'Dowd \& Urry 2005) and in the NIR (Cheung et al. 2003).}
The BAL shows dramatic variability in \zsh{its}  spectra
from the Sloan Digital Sky Survey (SDSS; York et al. 2000) \zsh{taken at five epochs},
\zsh{as well as from the previous spectra in the literature. }
We  present detailed spectral analysis in \S2, 
The results \zsh{are}  discussed in \S3, together
with future prospectives. Throughout this paper, we assume a
cosmology with $H_0= 70$ $\rm km~s^{-1}~Mpc^{-1}$, $\Omega_{\rm
M}=0.3$, and $\Omega_{\Lambda}=0.7$.

\section{\zsh{Broad absorption line and variability}}
\label{sec:spec}

\begin{figure}
  \begin{center}
  \includegraphics[angle=0, scale=0.7, trim= 6 10 5 10, clip]{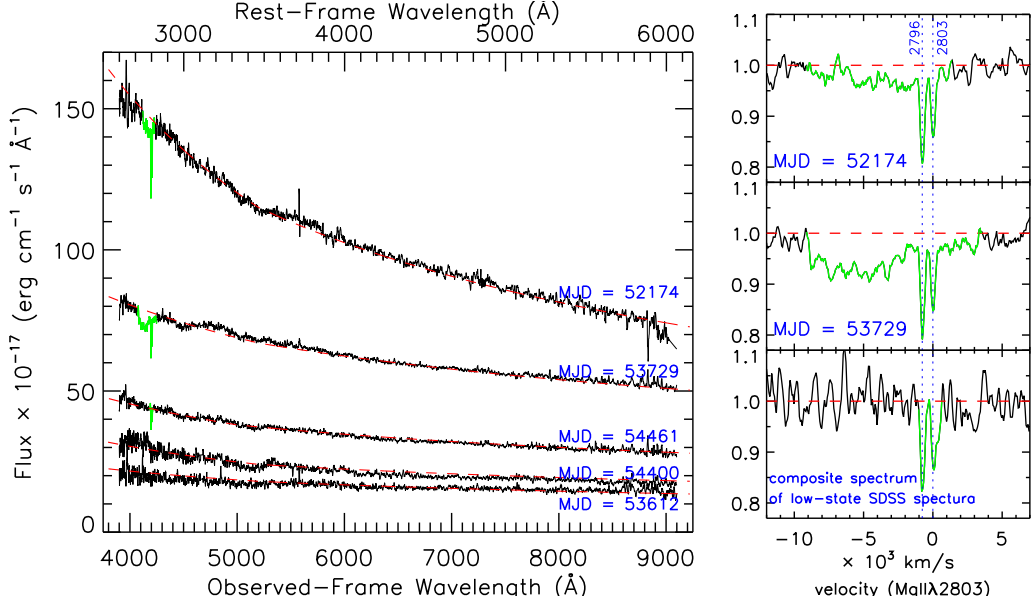} \\
  \end{center}
\caption{Left: The five SDSS optical spectra of \jn~. \zsh{The observed spectra from the SDSS are in back},
 we  plot the broken
power-law continuum in red, and absorption wavelength range in
green. The rest-frame wavelength is transformed from the
observed-frame wavelength using the absorption redshift
\mbox{$z_{\rm abs}=0.5$}. Right: The normalized spectra
obtained by dividing the optical spectra by the
broken power-law model continuum vs. velocity (relatived to
Mg~II$\lambda$2803\AA), the absorption structure in green and the
positions of Mg~II$\lambda\lambda$2796,2803 by blue dot
lines. The red short lines mean the value = 1.
\label{fig1} }
\end{figure}
The \zsh{Journals} of the five SDSS spectroscopy observations
\zsh{are} summarized in Table 1, together with \zsh{those} of three
historical spectroscopy observations collected in the literature. The SDSS
spectra are displayed in the left panel of Figure \ref{fig1}\zsh{, that is}
corrected for the Galactic reddening of E(B$-$V) = 0.029 mag
(Schlegel et al. 1998). \zsh{A} prominent feature is a BAL,
superimposed with a well-resolved Mg~II$\lambda
\lambda$2796.35,2803.53 doublet \zsh{of}  narrow absorption lines (NALs) at a
redshift of $z_{a} = 0.5006 \pm 0.0001$. The Mg~II doublet NALs are
obvious in MJD52174, 53729, and 54461\zsh{,} but can only be marginally
detected in MJD54400 and 53612 due to their low signal-to-noise
ratios (S/N)\footnote{We will denote the SDSS spectra using their
Modified Julian Date of observations, $MJD=JD-2400000$.}. The BAL
shows a dramatic variability\zsh{,} It is \zsh{detected} in the earliest
spectrum at the highest state, MJD52174\zsh{,} but \zsh{it disappeared} in MJD53612
at the lowest state; then \zsh{it appeared} itself again only about 100 days
later in MJD53729 at the second highest state with an enhanced
strength;  at last \zsh{disappeared} when the continuum \zsh{weakened} in
MJD54400 and 54461.

It should be pointed out that the BAL detected in \jn~ does not
strictly fulfill the conventional definitions, which require that
the absorption trough must fall at least
10\% below the model of continuum plus emission lines. Such a
conservative criterion was first suggested by Weymann et al. (1991;
see also Zhang et al. (2010) for a summary of BAL definitions) to
ensure \zsh{the} flux deficit is not \zsh{caused by} improper modeling \zsh{of the}
continuum and emission lines and/or low S/N. We note that
as a BL Lac \zsh{object}, the spectra of \jn~ are nearly featureless except the
absorption line regime, \zsh{that} greatly \zsh{simplifies the} recovering \zsh{of} the
absorption-free spectra. Besides, the high state SDSS spectra are of
rather high quality with a median $S/N\sim 50~\rm pixel^{-1}$. This
\zsh{enables} us to measure the BAL with much improved \zsh{significance} than in \zsh{the earlier}
quasar spectra. 
To further make sure that the BAL is  real, we extracted the raw 1-d spectra of all the
individual exposures of MJD52174 and 53729, without flux calibration
and rejection of the deviant pixels. The feature persists in all of
the individual spectra. This ensures that the BAL is not a false
feature introduced by the SDSS spectroscopic pipeline. We examined
the spectra of adjacent fibers, none of them \zsh{shows} a similar feature.
Besides, MJD52174 and 53729 \zsh{were} observed using different plates.
This excludes the possibility of the BAL \zsh{was} induced by either CCD
defect or improper CCD reduction. We also searched for spectroscopic
data in the literature, and found three historical spectra are
available (see Table 1). The BAL clearly shows itself in the
spectrum observed at a high state by Stickel et al. (1993; their
Figure 2, hereafter the reference is referred to as S93 and the
spectrum as LS86). The BAL \zsh{vanished} in the spectra at \zsh{a low state} by
Stocke \& Rector (1997; their Figure 3, hereafter the reference is
referred to as SR97 and the spectrum as MMT96) and Rector \& Stocke
(2001; their Figure 1, hereafter the reference as RS01 and the
spectrum as KP95).

The BAL and its variability pattern can be seen \zsh{clearly} in the
right panels of Figure \ref{fig1}, where the normalized SDSS
spectra around the BAL feature \zsh{are} plotted against velocity
relative to the Mg~II NALs. The continuum level is estimated by
fitting the spectra with a broken power law with the BAL masked.
We plot the two high state spectra individually, and combine the
three spectra at the low states to increase the S/N. We calculated
the observed frame equivalent width ($EW$) of the Mg~II NALs and the
BAL and listed them in Table 1. The strength of the Mg~II NALs
remains constant within the measurement errors in the 5 SDSS
spectra.
While the strength of the BAL changed dramatically from
$EW=6.35\pm 0.35$ \AA~in MJD53729 to $EW=1.91\pm 0.31$ \AA~in
MJD52174. Though the BAL strength \zsh{trebled} from MJD52174 to 53729,
the velocity structure largely remains the same, starting from
$-v_{min}\approx -3\times 10^{3}~\rm km~s^{-1}$ to a maximum
velocity of $-v_{max}\approx 10^{4}~\rm km~s^{-1}$.  We also
estimated the BAL strength in LS86 and found $EW\approx 3$ \AA,
which is between that in MJD52174 and 53729. Since there is no
significant BAL in MJD54461, 54400 and 53612 at \zsh{a low state}, we
derived an upper limit of $EW \lesssim 0.56$ \AA~(3 $\sigma$).
Moreover, the total observed frame EWs
estimated in KP95 and MMT96, are $\sim$1.1\AA, consistent with the
average $EW$ of Mg~II NAL. This suggests that there is no BAL in the
spectra of KP95 and MMT96, both of which have a continuum level
similar to the two epoch SDSS spectra at the lowest \zsh{state},
MJD53612 and 54400.

The identification of the BAL is not clear\zsh{, and } neither  the redshift
of the BL Lac \zsh{object}. The often quoted redshift of $z=0.733$ is probably
mis-estimated, that is based on detection of weak emission lines in
Mg~II$\lambda$2798, [Ne~V]$\lambda$3426 and
[O~II]$\lambda$3727, and \zsh{the} stellar absorption doublet
Ca~II$\lambda\lambda$3933,3968 in KP95 (RS01) and MMT96 (SR97).
Although we also detected the suspected weak signals of Mg~II and
[O~II] in the composite of the three epoch SDSS spectra
at \zsh{a} low state, this emission redshift is incongruous with the existence of \zsh{the} BAL.
The deep R-, K'-band imaging and \zsh{the} Hubble Space Telescope (HST)
images reveal four nearby non-stellar objects, including a
companion object within 1.44\arcsec~and three more fainter objects
within 5\arcsec~from this BL Lac object (Heidt et al. 1996; Scarpa et al.
2000). The emission lines detected by RS01 and SR97 possibly come
from the companion objects.
Plotkin et al. (2010) assigned a lower limit of $z\approx 0.501$
to \jn~ from the Mg~II doublet NALs.
In a recent paper, Bergeron et al. (2011) confirmed the excess of
intrinsic NAL in BL Lac objects using an enlarged sample of 38 BL Lac \zsh{objects}
including 3 candidates, and further reported a marginal excess
(1.5 $\sigma$) of associated/intrinsic absorbers with a velocity
relative to the BL Lacs of $\sim 0.1c$. We would be able to avoid
a chance coincidence for the Mg~II NALs and the unknown BAL in
\jn, provided that the Mg~II NALs \zsh{are} intrinsic to the BL Lac \zsh{object} and
the BAL is Mg~II. Otherwise, if the Mg~II NALs were an intervening
system, the BAL should be absorption \zsh{of ions} other than \zsh{Mg$^{+}$}.
Then we would have a rather small probability for the two observed
absorption features to coincide in wavelength in \jn. For
instance, if the BAL were a C~IV BAL, \zsh{that} is most commonly seen
in quasar spectra, we would have a post-probability of $p\lesssim
1\%$ for the Mg~II NALs occurring at the onset of the BAL. We
therefore \zsh{argue} that the Mg~II NALs are intrinsic to the BL Lac \zsh{object}
and the BAL is Mg~II BAL. We  adopt this assumption and  a
systematic redshift of $z\approx 0.5$ in the rest of this paper.
Note that most of the discussion in \S3 \zsh{is} not dependent on such
an assumption.

\begin{table}
\caption{Journals \zsh{of} the spectroscopic
observations and measured spectral parameters of \jn~ \label{tab1} }
\begin{center}
 \begin{tabular}{llcrrc}
 \hline
Spec.  &  Date-Obs.  & $F_{\rm 5000\AA}$  & $EW\rm^B$ & $EW\rm^N$ & references \\
\hline
LS86     & 1986/10          &101   &$\sim 3^a$             &  1.23 $^b$     &   S93 \\
KP95     & 1995/11/19,21    & 22   &                       & $\sim 1.1^c$   &   RS01\\
MMT96    & 1996/08/15       & 12   &                       & $\sim 1.1^c$   &   SR97\\
MJD52174 & 2001/09/22       &120   &$1.91\pm0.31 $         & $1.29\pm0.16~$ &       \\
MJD53612 & 2005/08/30       & 16   &$<$0.56$^d$            & $1.09\pm0.08^e$&       \\
MJD53729 & 2005/12/25       & 70   &$6.35\pm0.35 $         & $1.36\pm0.14~$ &       \\
MJD54400 & 2007/10/24       & 23   &$<$0.56$^d$            & $1.09\pm0.08^e$&       \\
MJD54461 & 2007/12/27       & 38   &$<$0.56$^d$            & $1.32\pm0.22~$ &       \\
\hline
\end{tabular}
\begin{list}{}{}
\item[ ] $EW\rm^B$ and $EW\rm^N$: the observed-frame equivalent widths of the broad and narrow absorption lines in units of \AA.
\item[ ] $F_{\rm 5000\AA}$: continuum levels in units of $\rm 10^{-17}~erg~s^{-1}~cm^{-2}~\AA^{-1}$.
\item[$^a$] $EW\rm^B$ is obtained by subtracted the average value of Mg~II NAL equivalent widths from the total absorption equivalent width estimated from the Figure 2 in S93.
\item[$^b$] the average value of Mg~II NAL equivalent widths obtained from SDSS spectra.
\item[$^c$] the absorption equivalent widths estimated from the Figure 3 in SR97 and Figure 1 in RS01.
\item[$^d$] $EW\rm^B$ is $3\sigma$ upper-limit estimated from spectrum fit for the low-state composite spectrum (MJD53612, 54400 and 54461)
using the broad absorption trough observed at MJD53729.
\item[$^e$] $EW\rm^N$ is measured from the composite spectrum of two low-state spectra (MJD53612 and 54400).
\end{list}
\end{center}
\end{table}

\section{Discussion and Summary}
\label{sec:discu}

BAL variability \zsh{sometimes}  seen in quasar spectra\zsh{, that} is useful for
exploring the geometrical and physical properties and evolution of
the BALR (e.g., Capellupo et al. 2010). The variability
is often attributed to a change in the ionization state and/or in CF
 of the BALR (e.g., Lundgren et al. 2007; Gibson et al. 2008).
Neither of the two precesses can naturally interpret what we
observed in \jn. On the one hand, we would need to \zsh{turn} the BALR
\zsh{on}  the three-epoch
spectra at \zsh{a high state} and to \zsh{turn it off}  in  the five-epoch
spectra at \zsh{a low state }for the CF change scenario. On the other hand,
photoionization timescales depend on both of the variability of the
energy input to the BALR and the density of the BALR. There is no
way to fine-tune the exterior and interior conditions to adapt well
to the observed BAL variability pattern. 
\zsh{Normal Mg~II BAL quasars tends to have high accretion rate,
while BL Lac objects are believed to have extremely low accretion rate.
These two kinds of  BAL AGNs may have different accretion modes,
and  further different origin and acceleration mechanism of BALR.}
Instead of \zsh{arising} from the accretion
disk as generally believed in quasars, the BAL outflow in \jn~ is
likely induced by the relativistic jet, \zsh{as we discussed in the following.}

\begin{figure}
  \begin{center}
\begin{tabular}{c}
\includegraphics[width=0.34\textwidth]{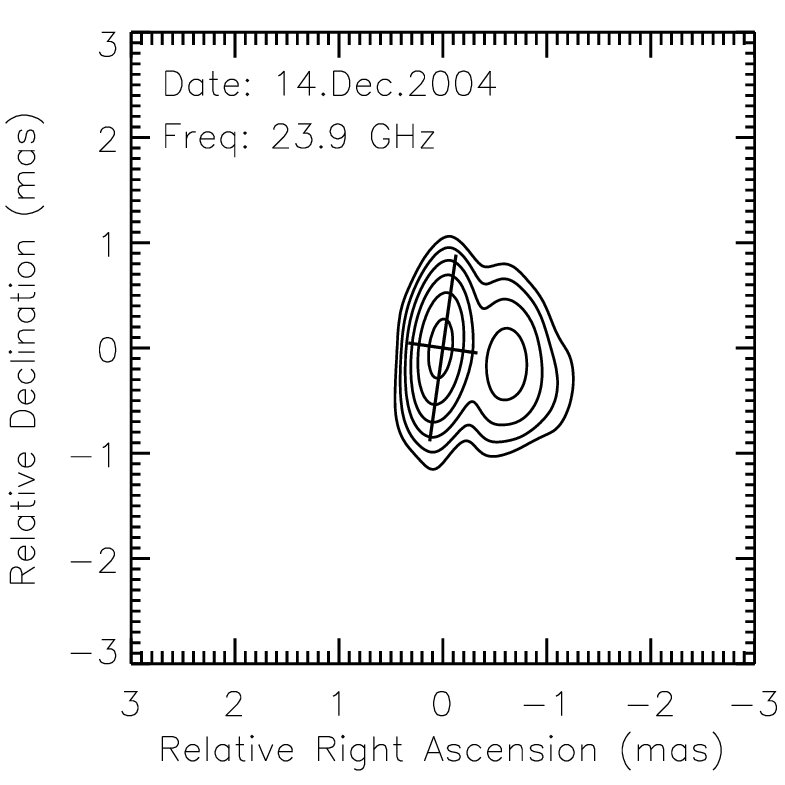}
\includegraphics[width=0.34\textwidth]{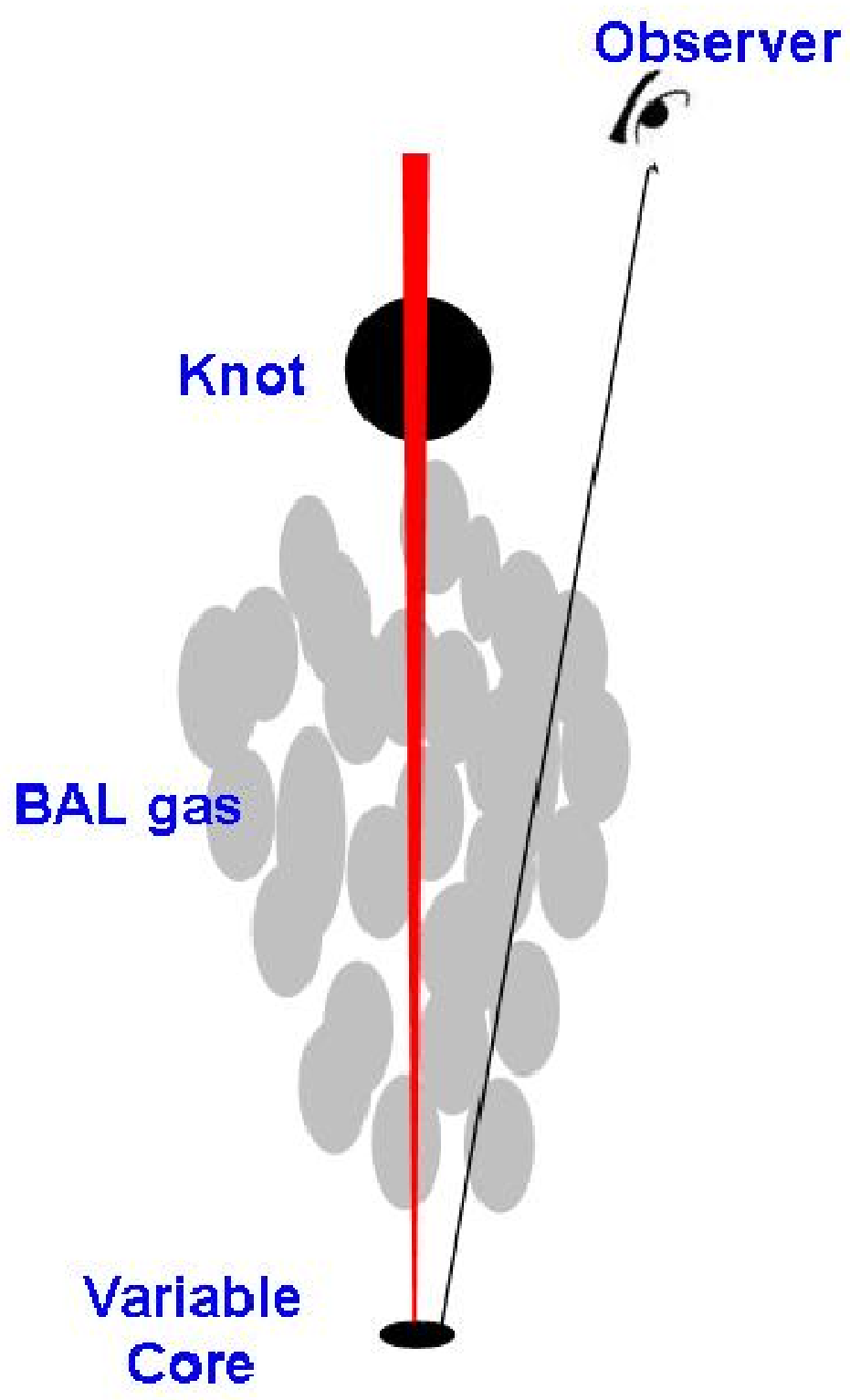}
\end{tabular}
  \end{center}
\caption{Left: The VLBA radio map at 23.9 GHz of \jn~. Peak fux=0.211 Jy/beam,
rms$\_$noise = 1.3 mJy/beam and the levels = 5, 10, 20, 40, 80, 161 mJy/beam.
Right: The edge-on viewing cartoon of the jet and BAL winds.
The gray clouds of the BAL winds are the gases entrained by the powerful jet.
Nested inside is a relativistic jet in red. The core and knot of jet are shown
by black clouds.
\label{fig2}}
\end{figure}

\jn~ has a core plus \zsh{an} one-sided jet of parsec scale as shown in the
high resolution radio map in Figure \ref{fig2} left panel(Fey et al. 2005).
The radio intensity ratio between the core and the jet is $\sim 2.68$.
In the standard AGN unification schemes, BL Lac \zsh{objects} are described as
Fanaroff Riley type I Radio Galaxies (FR-I~RGs) with their relativistic jets pointed
toward the observer (e.g., Blandford \& Rees 1978; Urry \& Padovani 1995).
We may use M87, the adjacent FR-I RG with well-resolved jet in
multi-wavelength, as a guide line in our discussion. The radio
through optical spectral index is $\alpha_{ro}\sim 0.70$ for the
core and $\alpha_{ro}\sim 0.69$ for the jet ($S_{\nu}\propto
\nu^{-\alpha_{ro}}$; e.g., Wang \& Zhou 2009; see also Meng \&
Zhou 2006 for more examples). Assuming these values, we obtained a
rough estimate of the optical intensity ratio of $\sim$ 2.22
between the core and the jet in \jn. This suggests that both of
the core and the jet may make significant contribution to the \zsh{observed}
optical continuum emission. It is easier for the BALR to obscure
the core than to obscure the jet, since the jet is much more
extended than the core. Instead of the variability of the CF or
the ionization level of the BALR, the unusual BAL variability
pattern observed in \jn~ likely \zsh{results} from the variable ratio
between the emission \zsh{components} from the core and the jet. We create a
cartoon in Figure \ref{fig2} \zsh{to illustrate} this simple
scenario. The right panel shows the edge-on viewing of the jet and
BAL winds. The BAL material is located somewhere between the core
and the knot. As the optical emission is dominated by the core,
BAL feature appears\zsh{, otherwise the} BAL disappears.

It has been suggested that jet-cloud interaction is a common
phenomenon in Blazars  at both
small and large scales (Jones 1996; Yuan et al.2000; G{\'o}mez et al. 2006;
Araudo et al. 2009; Reynolds et al. 2009; Bergeron et al. 2011). This
process may serve as a natural mechanism to accelerate gas to high
velocity. That is \zsh{possibly the }reason for \zsh{the} BAL occurrence in BL Lac \zsh{objects}
with extremely low accretion \zsh{rates}. It is unclear the frequency of \zsh{the}
BAL occurrence in BL Lac \zsh{objects}, or its dependence on the BL Lac properties.
We examine the optical and near-IR
spectra of 20 objects with emission or absorption redshift $ min
\{[z_{\rm emi}, z_{\rm abs}]\} \ge 0.4$ in all the 34 BL Lac \zsh{objects}
\zsh{objects in the well defined 1 Jy sample}, but
we do not find \zsh{any other  BL Lac objects showing BAL}. The fraction of BAL
occurrence in BL Lac \zsh{objects} is $\sim5\%$ in \zsh{the} 1 Jy sample. Considering the
\zsh{non-detetion} of BAL because of the low spectrum S/N and strong
BAL variability, such as \jn, the true fraction of BAL BL Lacs is
likely to be higher.

In this Letter, we report the first detection of a BAL in a BL
Lac \zsh{object}.
\zsh{The concurrence of Mg~II~NALs and the BAL suggested that
the intrinsic redshift of this object is $z \sim 0.5$ and the BAL is Mg~II~BAL.
However these results are tentative and further confirmation is needed.
The Mg~II~BAL appeared with a similar velocity structure in the two epoch optical spectra
at a high state, and disappeared  in all the three } low state spectra. The BAL
material is \zsh{suggested} to possibly locate somewhere between the
core and the knot, \zsh{and} is accelerated as the jet propagates through
it. The BAL variability \zsh{is} likely caused by the \zsh{varying} relative
contribution of \zsh{the} continuum emission from the core and the jet.
\zsh{This scenario can be confirmed by detection of
the expected correlation between the core emission and
the BAL strength via high resolution radio mapping and
optical spectrophotometric monitoring, respectively.}

\begin{acknowledgements}
\zsh{We thank the anonymous referee for helpful comments.}
This work was supported by the China NSF grant 10973012 and 11033007,
and the National Basic Research Program of China (973 program, 2007CB815405
and 2009CB824800).

Funding for the SDSS and SDSS-II has been provided by the Alfred P.
Sloan Foundation, the Participating Institutions, the National
Science Foundation, the U.S. Department of Energy, the National
Aeronautics and Space Administration, the Japanese Monbukagakusho,
the Max Planck Society, and the Higher Education Funding Council for
England. The SDSS Web site is http://www.sdss.org/.
\end{acknowledgements}

\end{document}